\documentclass[fleqn, 12pt] {article}
\usepackage{epsf}
\begin{document}
\title{Combined effect of viscosity and vorticity on single mode Rayleigh -Taylor instability bubble growth}
\author{Rahul Banerjee\thanks{e-mail: rbanerjee.math@gmail.com}, Labakanta Mandal\thanks{e-mail:
labakanta@gmail.com}, S. Roy\thanks{e-mail:
phy.sou82@gmail.com}, M. Khan\thanks{e-mail: mkhan$_{-}$ju@yahoo.com}, M. R. Gupta\thanks{e-mail: mrgupta$_{-}$cps@yahoo.co.in} \\
Deptt. of Instrumentation Science \& Centre for Plasma Studies
\\Jadavpur University, Kolkata-700032, India\\}
\date{}

\maketitle
\begin{abstract}
The combined effect of viscosity and vorticity on the growth rate
of the bubble associated with single mode  Rayleigh -Taylor
instability is investigated. It is shown that the effect of
viscosity on the motion of the lighter fluid associated with
vorticity accumulated inside the bubble due to mass ablation may
be such as to reduce the net viscous drag on the bubble exerted by
the upper heavier fluid as the former rises through it.
\end{abstract}
\newpage

\section*{I INTRODUCTION}
Recently obtained simulation results$^{\cite{pr06}}$ regarding
single mode Rayleigh Taylor instability(RTI) problem indicates
that the bubble developed at the two fluid interface is
accelerated to a velocity quite above the classical model
value$^{\cite{vg02}}$ . This is ascribed to vorticity formation on
the bubble-spike interface that is suggested$^{\cite{pr06}}$  to
diminish the friction drag due to viscosity. The theoretical
calculations and the consequent numerical results regarding
Rayleigh Taylor instability in a viscous fluid$^{\cite{for09}}$
also indicates growth of vorticity inside the bubble in the
neighbourhood of its tip, where the curvature is maximum. Recent
experimental and simulation results$^{\cite{hsp10}}$ show that the
influence of viscosity on RTI may strongly suppress the growth
rate. It is suggested that high pressure and strain rate
conditions give rise to a phonon drag mechanism resulting in the
lattice viscosity provided a solid state is
maintained$^{\cite{hsp10}}$. Moreover, it was also suggested by
Betti and Sanz$^{\cite{bt06}}$ that at the spike, mass ablation
induces a transverse component of velocity and thus gives rise to
vorticity generation. It has been shown by the same authors that
the instability growth rate is augmented by vorticity accumulation
inside the bubble resulting from mass ablation.

As the bubble rises through the denser fluid it encounters viscous
resistance. In this paper we have incorporated the joint effect of
viscosity and vorticity. It is shown how the vorticity induced
contribution to bubble velocity as derived by Betti and
Sanz$^{\cite{bt06}}$ is affected by the viscous drag exerted by
the lighter fluid inside the bubble. It is found that the combined
effect of vorticity and viscosity of the bubble fluid tends to
diminish the friction drag exerted by the upper heavier fluid as
the bubble penetrates through it. In this respect it is similar to
the friction drag reduction effect due to vorticity as suggested
by the phenomenological model of Ramaprabhu et al.$^{\cite{pr06}}$
.

Section II contains the mathematical formulation as also the two
fluid boundary conditions. The results and discussions are
presented is section III.

\section*{II MATHEMATICAL FORMULATION AND INTERFACIAL BOUNDARY CONDITION}

We are considering 2D single mode Rayleigh-Taylor instability
problem. The $y$ axis is taken vertically upward opposite to the
direction of gravity and $x$ axis taken along the horizontal plane
of unperturbed two fluid interface plane. The equation to the
perturbed surface of bubble which is assumed to be parabolic is
\begin{eqnarray}\label{eq:1}
y\equiv\eta(x,t)=\eta_{0}(t)+\eta_{2}(t)x^2; \quad y>0
\end{eqnarray}
with $\eta_{0}(t)>0$ and $\eta_{2}(t)<0$ (terms
O($x^{2+i}$)($i\geq1$) are neglected as we are interested only in
the motion very close to the tip of the bubble
$|x|\ll1$)$^{\cite{mrg10}}$.

The kinematic boundary conditions satisfied on the perturbed
interface given by Eq.(1) are
\begin{eqnarray}\label{eq:2}
\frac{\partial\eta}{\partial t}+v_{hx}
\frac{\partial\eta}{\partial x}=v_{hy}
\end{eqnarray}
\begin{eqnarray}\label{eq:3}
\frac{\partial\eta}{\partial x}(v_{hx}-v_{lx})=v_{hy}-v_{ly}
\end{eqnarray}
where $(v_{h,l})_{x,y}$ are the velocity components of the heavier
(density $\rho_{h}$) and lighter (density $\rho_{l}$) fluids
occupying the regions $y>0$ and $y<0$ respectively.

Assume single mode potential flow for the upper fluid governed by
the potential
\begin{eqnarray}\label{eq:4}
\phi_h(x,y,t)=a(t) \cos(kx) e^{-k(y-\eta_{0}(t))}
\end{eqnarray}
with $v_{hx}=-\frac{\partial\phi_{h}}{\partial x}$ and
$v_{hy}=-\frac{\partial\phi_{h}}{\partial y}$

Since both $\nabla^2 v_{hx}=0$ and $\nabla^2 v_{hy}=0$ for
potential flow, the equation of motion of the upper incompressible
fluid of constant density $\rho_h$ and coefficient of kinematic
viscosity $\nu_{h}$ ($\mu_{h}=\rho_{h}\nu_{h}$)

\begin{eqnarray}\label{eq:5}
\rho_{h}[\frac{\partial \vec{v_{h}}}{\partial
t}+(\vec{v_{h}}.\vec{\nabla})\vec{v_{h}}]+\vec{\nabla}(p_{h}+g\rho_{h}y)+\mu_{h}\nabla^2\vec{v_{h}}=\vec{0}
\end{eqnarray}
leads to the following integral for the irrotational motion.
\begin{eqnarray}\label{eq:6}
\rho_{h}[-\frac{\partial \phi_{h}}{\partial t}+
\frac{1}{2}(\vec{\nabla} \phi_{h})^{2}+ g y]+p_{h}=f_{h}(t)
\end{eqnarray}
For the lighter density fluid the flow inside the bubble is
assumed rotational$^{\cite{bt06}}$ with vorticity
$\vec{\omega}=(\frac{\partial v_{ly}}{\partial x}-\frac{\partial
v_{lx}}{\partial y}){\hat{z}}$. The motion is described by the
stream function $\Psi(x,y,t)$;
\begin{eqnarray}\label{eq:7}
v_{lx}=-\frac{\partial \Psi}{\partial y} \mbox{ and }
v_{ly}=\frac{\partial \Psi}{\partial x}
\end{eqnarray}
so that
\begin{eqnarray}\label{eq:8}
\nabla^2 \Psi =-\omega
\end{eqnarray}
Let $\chi(x,y,t)$ be a function such that
\begin{eqnarray}\label{eq:9}
\nabla^2 \chi =-\omega
\end{eqnarray}
Hence $(\Psi-\chi)$ is a harmonic function as $\nabla^2
(\Psi-\chi)=0$. Let $\Phi(x,y,t)$ be its conjugate function
\begin{eqnarray}\nonumber
\frac{\partial \Phi}{\partial x}=\frac{\partial \Psi}{\partial
y}-\frac{\partial \chi}{\partial y}
\end{eqnarray}
\begin{eqnarray}\label{eq:10}
\frac{\partial \Phi}{\partial y}=-\frac{\partial \Psi}{\partial
x}+\frac{\partial \chi}{\partial x}
\end{eqnarray}
i.e. that velocity components of the lighter fluid are
\begin{eqnarray}\nonumber
v_{lx}=-\frac{\partial \Psi}{\partial y}=-\frac{\partial
\Phi}{\partial x}-\frac{\partial \chi}{\partial y}
\end{eqnarray}
\begin{eqnarray}\label{eq:11}
v_{ly}=\frac{\partial \Psi}{\partial x}=-\frac{\partial
\Phi}{\partial y}+\frac{\partial \chi}{\partial x}
\end{eqnarray}
The following choice is made for the stream function according to
$^{\cite{bt06}}$
\begin{eqnarray}\label{eq:12}
\Psi(x,y,t)=b_{0}(t)x+[b_{1}(t)e^{k(y-\eta_{0})}+\omega_{0}(t)/k^2]
\sin{(k x)}
\end{eqnarray}
with
\begin{eqnarray}\label{eq:13}
\chi(x,y,t)=\omega_{0}(t)\sin{(k x)}/k^2
\end{eqnarray}
Eq.(10) now gives
\begin{eqnarray}\label{eq:14}
\Phi(x,y,t)=-b_0(t)y+b_1(t)\cos{(k x)} e^{k(y-\eta_0)}
\end{eqnarray}
Substituting for the velocity components as determined by the
velocity potential $\phi_{h}$ and the stream function $\Psi$ in
the kinematic boundary conditions Eq.(1) and Eq.(2) and equating
coefficients of $x^i$ for $i=0$ and $2$ we obtain the following
four equations for nondimensionalized bubble height $
\xi_{1}=k\eta_{0}$, nondimensionalized curvature
$\xi_{2}=\eta_{2}/ k$ and $b_{0}$, $b_{1}$
\begin{eqnarray}\label{eq:15}
\frac{d\xi_{1}}{d\tau }=\xi_{3}
\end{eqnarray}
\begin{eqnarray}\label{eq:16}
\frac{d\xi_{2}}{d\tau}=-\frac{1}{2}(6\xi_{2}+1)\xi_{3}
\end{eqnarray}
\begin{eqnarray}\label{eq:17}
\frac{kb_{0}}{\sqrt{kg}}=\frac{6\xi_{2}(2\xi_{3}-\hat{\omega})}{(6\xi_{2}-1)}
\end{eqnarray}
\begin{eqnarray}\label{eq:18}
\frac{k^2b_{1}}{\sqrt{kg}}=-\frac{(6\xi_{2}+1)\xi_{3}-\hat{\omega}}{(6\xi_{2}-1)}
\end{eqnarray}
Here
\begin{eqnarray}\label{eq:19}
\xi_{3}=\frac{k^2a}{\sqrt{kg}}
\end{eqnarray}
is the nondimensionalized velocity of the tip of the bubble,
\begin{eqnarray}\label{eq:20}
\tau=t\sqrt{kg}
\end{eqnarray}
is the nondimensionalized time and
\begin{eqnarray}\label{eq:21}
\hat{\omega}=\frac{\omega_{0}}{\sqrt{kg}}
\end{eqnarray}
is the nondimensionalized vorticity.

  We need one more equation, i,e., the one for $\xi_{3}(\tau)$ to complete the set of five necessary
equations for the five functions describing the time evolution of
bubble. This is obtained from the dynamical boundary condition as
done below.

Starting from the equation of motion of the lighter (lower) fluid
with coefficient of kinematic viscosity $\nu_l$
($\mu_l=\rho_l\nu_l$)
\begin{eqnarray}\label{eq:22}
\rho_{l}[\frac{\partial \vec{v}_{l}}{\partial t}+
\frac{1}{2}\vec{\nabla}(\vec{v}_{l} {^2})+ \omega \hat{z}\times
\vec{v}_{l}]+ \vec{\nabla} (p_{l} + \rho_{l} g y)-\mu_{l} \nabla^2
\vec{v}_{l}=0
\end{eqnarray}
and using Eqs.(7)-(11) we get
\begin{eqnarray}\nonumber
\rho_{l}[-\frac{\partial \Phi}{\partial t}+
\frac{1}{2}(\vec{v}_{l} ){^2}-\omega \Psi +
\frac{p_{l}}{\rho_{l}}+g y]+\int \rho_{l}[(\Psi \frac{\partial
\omega}{\partial y}-\frac{\partial \dot{\chi}}{\partial y})dx
+(\Psi \frac{\partial \omega}{\partial x}+\frac{\partial
\dot{\chi}}{\partial x})dy]
\end{eqnarray}
\begin{eqnarray}\label{eq:23}
+\int\mu_{l}(\frac{\partial \omega}{\partial y}dx -\frac{\partial
\omega}{\partial x}dy)=0
\end{eqnarray}
The last term on the L.H.S being the contribution from viscous
drag.

For rotational motion Bernoulli's equation does not exist except
for special cases$^{\cite{sfmn92}}$. However for velocity
potential and stream function as given by Eqs.(12)-(14) the
integrations in Eq.(23) can be expressed in closed
form$^{\cite{bt06}}$.

Taking the difference of Eq.(6) and Eq.(23) at the interface of
the two fluids and applying the dynamical boundary condition, i.e,
the pressure difference at the interface is balanced by the
difference of viscous stress and surface tension$^{\cite{ss03}}$
we obtain the following:
\begin{eqnarray}\nonumber
\rho_{h}[-\frac{\partial \phi_{h}}{\partial t}+
\frac{1}{2}(\vec{\nabla} \phi_{h})^{2}+ g
y]-\rho_{l}[-\frac{\partial \Phi}{\partial t}+
\frac{1}{2}(\vec{\nabla}\Phi ){^2}-\omega \Psi +g y]-\int
\rho_{l}[(\Psi \frac{\partial \omega}{\partial y}-\frac{\partial
\dot{\chi}}{\partial y})dx +(\Psi \frac{\partial \omega}{\partial
x}+\frac{\partial \dot{\chi}}{\partial x})dy]
\end{eqnarray}
\begin{eqnarray}\nonumber
-\int\mu_{l}(\frac{\partial \omega}{\partial y}dx -\frac{\partial
\omega}{\partial x}dy)=-[p_{h}-p_{l}]+\tilde{f}(t)
\end{eqnarray}
\begin{eqnarray}\nonumber
=(2\mu_{h}\frac{\partial v_{hy}}{\partial
y}-2\mu_{l}\frac{\partial v_{ly}}{\partial
y})+\frac{T}{R}+\tilde{f}(t)
\end{eqnarray}
\begin{eqnarray}\label{eq:24}
 =-2\mu_{h}\frac{\partial^2 \phi_{h}}{\partial
y^2}+2\mu_{l}\frac{\partial^2 \Psi }{\partial x
\partial y}+\frac{T}{R}+\tilde{f}(t)
\end{eqnarray}
where $T$ is the surface tension and $R$ is the radius of
curvature at the interface.

Substituting for $\phi_{h}, \Psi, \chi,\Phi$, using Layzer's
approximation method$^{\cite{dl55}}$, expanding in power of x to
O($x^{i}$)($i\leq2$) and equating coefficient of $x^2$ we obtain
after some laborious but straightforward manipulation the time
dependence equation for $\xi_{3}(\tau)$. The set of time evolution
equations of the bubble are thus given by [together with Eq.(17)
and Eq.(18)] the following three differential equations for
$\xi_{1}, \xi_{2}$ and $\xi_{3}$
\begin{eqnarray}\label{eq:25}
\frac{d\xi_1}{d\tau}=\xi_3
\end{eqnarray}
\begin{eqnarray}\label{eq:26}
\frac{d\xi_2}{d\tau}=-\frac{1}{2}\xi_3(6\xi_2 + 1)
\end{eqnarray}
\begin{eqnarray}\nonumber
\frac{d\xi_3}{d\tau}=[-N(\xi_2,r)\frac{\xi_3^2}{(6\xi_2-1)}+2(r-1)(6\xi_2-1)\xi_2(1-12\xi_{2}^2\frac{k^2}{k_{c}^2})]\frac{1}{D(\xi_2,r)}
\end{eqnarray}
\begin{eqnarray}\nonumber
+[\frac{\hat{\omega}^2-5(6\xi_{2}+1)\hat{\omega}\xi_3}{(1-6\xi_2)}+\dot{\hat{\omega}}]\frac{1}{D(\xi_2,r)}
\end{eqnarray}
\begin{eqnarray}\label{eq:27}
-c_{h}r[(s+1)(1-12\xi_{2}^2)+4\xi_{2}(s-1)]\frac{2\xi_3}{D(\xi_2,r)}+\frac{2c_{h}
r s \hat{\omega}(1+2\xi_2)(1-3\xi_2)}{D(\xi_2,r)}
\end{eqnarray}
where
\begin{eqnarray}\label{eq:28}
N(\xi_2,r)=36(1-r)\xi_{2}^{2}+12(4+r)\xi_{2}+(7-r)
\end{eqnarray}
\begin{eqnarray}\label{eq:29}
D(\xi_2,r)=12(1-r)\xi_{2}^{2}+4(1-r)\xi_{2}+(r+1)
\end{eqnarray}
\begin{eqnarray}\label{eq:30}
k_c^2=\frac{(\rho_h-\rho_l)g}{T}; \mbox{ } s=\frac{\mu_l}{\mu_h};
\mbox{ } \nu_h=\frac{\mu_h}{\rho_h}; \mbox{ }
r=\frac{\rho_h}{\rho_l}; \mbox{ } c_h=\frac{\nu_h k^2}{\sqrt{kg}}
\end{eqnarray}

\section*{III NUMERICAL RESULTS AND DISCUSSIONS  }

The time evolution of bubble is to be worked out by numerical
integration of Eqs.(25)-(27) by employing Runge-Kutta-Fehlberg
scheme. To this end it is necessary to know the dependence of the
vorticity $\hat{\omega}(\tau)$ on $\tau$. We choose
$\hat{\omega}(\tau)$ in the following form so as to have a time
dependence having close resemblance with the simulation
results$^{\cite{bt06}}$.
 \begin{eqnarray}\label{eq:31}
\hat{\omega}(\tau)=\frac{\hat{\omega}_{c}}{1+2\tanh(\tau_{0})}[\tanh(\tau_{0})(1+\tanh(\tau))+\tanh(\tau-\tau_{0})]
\end{eqnarray}
Note that $\hat{\omega}(\tau)$ increases from $\hat{\omega}(0)=0$
and gradually builds up to a saturation value $\hat{\omega}_{c}$.
The constants $\tau_{0}$ and $\hat{\omega}_{c}$ are adjusted so
that the time dependence has approximate qualitative agrement with
simulation results given by Fig.4 of Ref.$^{\cite{bt06}}$. This
may be seen from the graph of $\hat{\omega}(\tau)$ plotted against
$\tau$ in Fig.1.

In Fig.2 are shown the nondimensionalized height ($\xi_{1}$),
curvature ($\xi_{2}$) and velocity ($\xi_{3}$) of the tip of the
bubble and plotted against time $\tau=\sqrt{kg}$.

(a). The first two terms
$[-N(\xi_2,r)\frac{\xi_3^2}{(6\xi_2-1)}+2(r-1)(6\xi_2-1)\xi_2(1-12\xi_{2}^2\frac{k^2}{k_{c}^2})]\frac{1}{D(\xi_2,r)}$
give the classical bubble velocity based on Goncharov
model$^{\cite{vg02}}$ modified by effect of surface tension.

(b). The third term
$[\frac{\hat{\omega}^2-5(6\xi_{2}+1)\hat{\omega}\xi_3}{(1-6\xi_2)}+\dot{\hat{\omega}}]\frac{1}{D(\xi_2,r)}
$ is the contribution from vorticity accumulation inside the
bubble$^{\cite{bt06}}$ causing enhancement of the bubble velocity.

(c). The fourth term
$-2c_{h}r[(s+1)(1-12\xi_{2}^2)+4\xi_{2}(s-1)]\frac{\xi_{3}}{D(\xi_2,r)}$
accounts for the lowering of the bubble velocity due to viscous
drag exerted  by the denser fluid.

(d). The last (fifth) term $2c_{h} r s
\hat{\omega}(1+2\xi_2)(1-3\xi_2)/ D(\xi_2,r)$ gives the vorticity
dependent contribution that diminishes the friction drag; such a
possibility was suggested by Ramaprabhu et al.$^{\cite{pr06}}$ .

Eq.(26) and Eq.(27) show that as $\tau\rightarrow\infty$ the
bubble velocity reaches the following (nondimensionalized)
asymptotic steady value with $\xi_{2}\rightarrow -\frac{1}{6}$
($A$= Atwood number),
$\hat{\omega}(\tau\rightarrow\infty)=\hat{\omega}_{c}$ and
$\dot{\hat{\omega}}(\tau\rightarrow\infty)=0$,

\begin{eqnarray}\label{eq:32}
(v_{b})_{asymp}=\sqrt{\frac{2}{3}\frac{A}{1+A}+\frac{\hat{\omega}_{c}^2}{4}\frac{1-A}{1+A}+\frac{4}{9}c_{h}^2
+sc_{h}\hat{\omega}_{c}}-\frac{2}{3}c_{h}
\end{eqnarray}

It is be noted that this result agrees with the saturation value
of  Betti and Sanz$^{\cite{bt06}}$ when viscosity is
neglected($c_{h}=0$) and with that of Sung-Ik-Sohn$^{\cite{ss03}}$
when vorticity is neglected.

 In Fig.2 the physical process(a)
is represented by plot I. It is the Goncharov (classical) model
result
\begin{eqnarray}\label{eq:33}
(v_{b})_{asymp}^{Gon}=\sqrt{\frac{2}{3}\frac{A}{1+A}}
\end{eqnarray}

In absence of viscosity the vorticity aided time development of
the bubble velocity is given by plot II in Fig.2. This represents
the combined effect (a)+(b). As $\hat{\omega}(\tau)$ increases
gradually from $\hat{\omega}(\tau=0)=0$, plot II indicates that so
does the bubble velocity $\xi_{3}(\tau)$ from the initial value
$\xi_{3}(\tau=0)$ remaining greater than the
$[\xi_{3}(\tau)]_{classical}$ but close to it as long as the
vorticity does not increase appreciably. This continues as $\tau$
increases till $\hat{\omega}(\tau)$ builds up sufficiently as
indicated in Fig.1 with a consequent rapid growth of
$\xi_{3}(\tau)$ above the classical saturation value
$[\xi_{3}(\tau)]_{classical}$ as given in plot I of Fig.2. As
$\tau\rightarrow\infty$ the vorticity aided bubble velocity
approches the asymptotic value
\begin{eqnarray}\label{eq:34}
(v_{b})_{asymp}^{rot}=\sqrt{\frac{2}{3}\frac{A}{1+A}+\frac{\hat{\omega}_{c}^2}{4}\frac{1-A}{1+A}}
\end{eqnarray}

As the bubble rises the viscous resistance of the upper(heavier)
fluid depresses the vorticity aided bubble velocity. The magnitude
of this depression depends on $c_{h}=\rho_{h}\nu_{h}$ . This is
shown by plot III in Fig.2 representing the cumulative effect
(a)+(b)+(c). During the initial stage, Plot III is very close to
plot I ($[\xi_{3}(\tau)]_{classical}$) due to vorticity and
depressing effect of viscosity. But as $\hat{\omega}(\tau)$
becomes sufficiently large as $\tau$ increases plot III shoots
above plot I and attain the asymptotic value
\begin{eqnarray}\label{eq:35}
(v_{b
}^{vis})_{asymp}^{rot}=\sqrt{\frac{2}{3}\frac{A}{1+A}+\frac{\hat{\omega}_{c}^2}{4}\frac{1-A}{1+A}+\frac{4}{9}c_{h}^2
}-\frac{2}{3}c_{h}
\end{eqnarray}

However due to accumulation of vorticity inside the bubble the
viscous drag is reduced. This is indicated by the presence of the
term $sc_{h}\hat{\omega}_{c}$. The possibility of such an
eventuality was suggested by Ramaprabhu et. al.$^{\cite{pr06}}$
and this is exhibited by the last plot IV in Fig.2 representing
the cumulative effect (a)+(b)+(c)+(d) which lies above plot III
but below plot II. Thus the following inequalities hold among the
asymptotic values:
\begin{eqnarray}\label{eq:36}
(v_{b }^{vis})_{asymp}^{rot}<
(v_{b})_{asymp}<(v_{b})_{asymp}^{rot}
\end{eqnarray}
The leftmost inequality shows that the viscous drag exerted by the
denser fluid as the bubble penetrates through it is reduced by the
viscosity effect on the motion of the lighter fluid due to
vorticity accumulation inside the bubble.

\section*{ACKNOWLEDGMENTS }
Rahul Banerjee acknowledges the support of C.S.I.R, Government of
India under ref. no. R-10/B/1/09.

\begin{figure}[p]
\vbox{\hskip 1.cm \epsfxsize=12cm \epsfbox{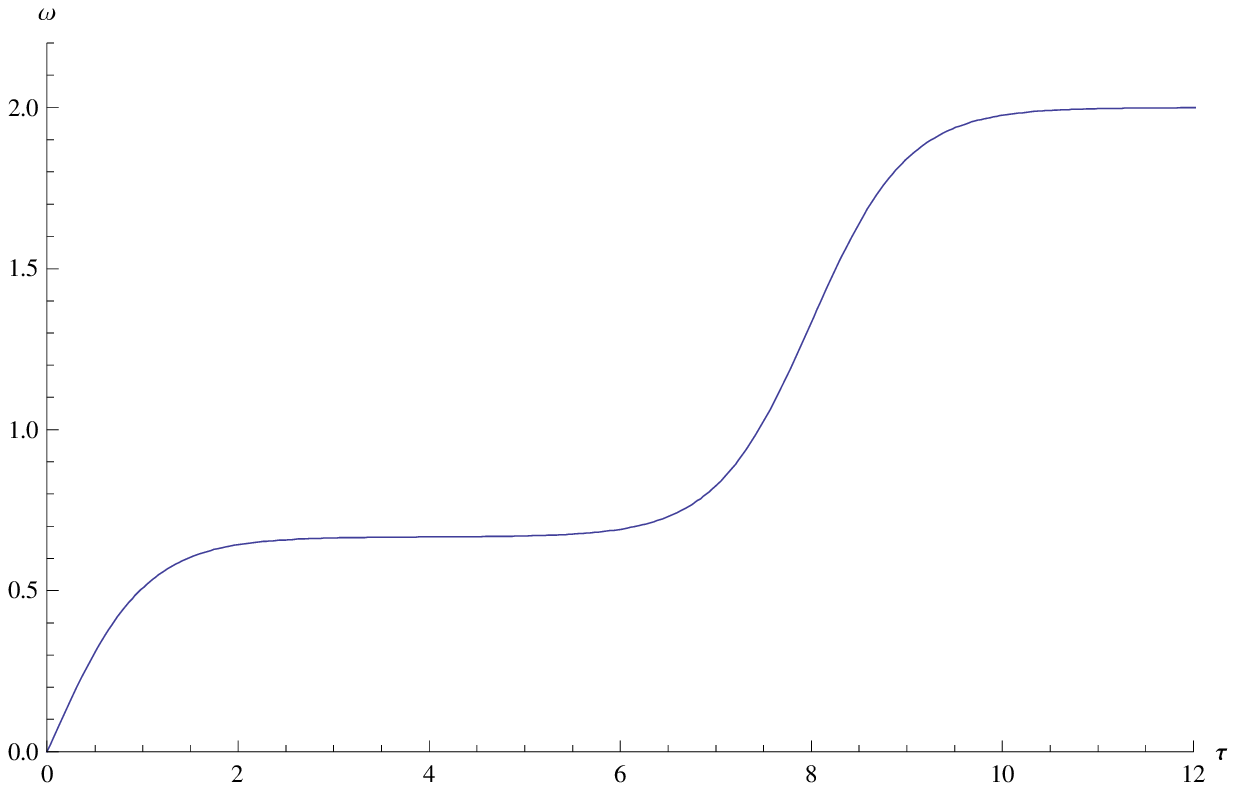}}
\begin{verse}
\vspace{-0.1cm} \caption{Vorticity $\hat{\omega}(\tau)$ plotted
against $\tau$ with saturation value $\hat{\omega}_{c}$=2 and
parameter $\tau_{0}$=8. }\label{Fig:1}
\end{verse}
\end{figure}

\begin{figure}[p]
\vbox{\hskip 1.cm \epsfxsize=12cm \epsfbox{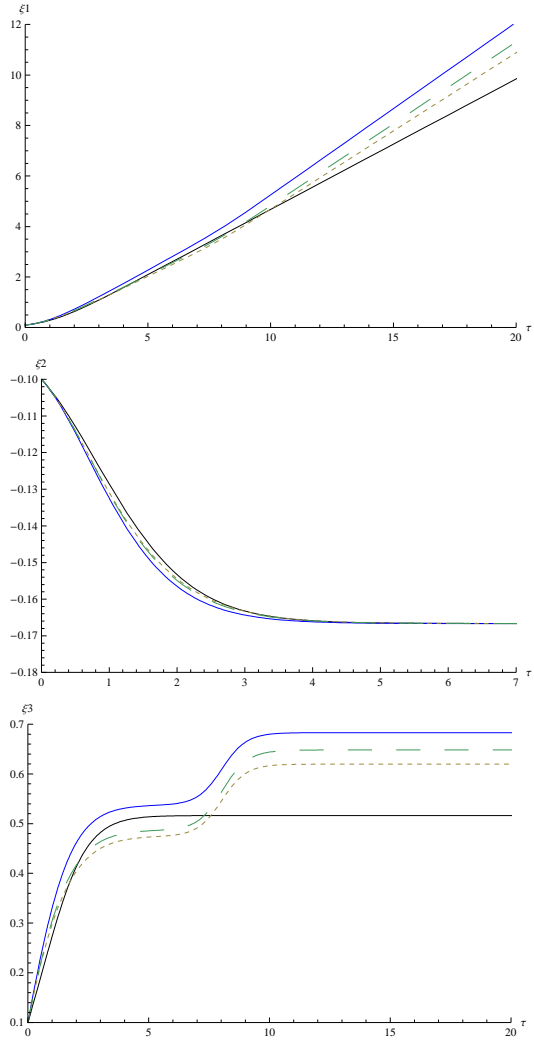}}
\begin{verse}
\vspace{-0.1cm} \caption{$r=5$, $s=1/5$, $\hat{\omega}_{c}=2$,
initial value $\xi_1=-\xi_2=\xi_3=0.1$. Black line
($\hat{\omega}(\tau)=0$, $c_{h}=0$), blue line
($\hat{\omega}(\tau)$ in Fig.1, $c_{h}=0$), dotted line
($\hat{\omega}(\tau)$ in Fig.1, $c_{h}=0.2$) and dash line
($\hat{\omega}(\tau)$ in Fig.1, $c_{h}=0.2$) represents plot I,
plot II, plot III and plot IV respectively.}\label{Fig:2}
\end{verse}
\end{figure}

\end{document}